\documentclass[prb,twocolumn,showpacs,preprintnumbers,amsmath,amssymb]{revtex4}

\usepackage[dvips]{graphicx}

\usepackage{amssymb}

\usepackage{bm}

\usepackage{amsmath}

\usepackage{color}

\begin{document}


\title{Interacting spin-droplets and magnetic properties of a low-density two-dimensional electron gas}

\author{Yevgeny V. Stadnik}
\affiliation{School of Physics, University of New South Wales, Sydney 2052, Australia}
\author{Oleg P. Sushkov}
\affiliation{School of Physics, University of New South Wales, Sydney 2052, Australia}

\date{\today}

\begin{abstract}

We argue that the magnetic susceptibility data, Refs. \cite{Reznikov_09,Reznikov_10,Reznikov_12}, for the low-density two-dimensional (2D) silicon-based electron gas indicate that magnetically active electrons are localised in spin-droplets. The droplets exist in both the insulating and metallic phases, and interact ferromagnetically, forming an effective 2D Heisenberg ferromagnet. Comparing the data with known analytical and numerical results for a 2D Heisenberg ferromagnet, we determine that $JS^2 \approx 0.6~\textrm{K}$, where $S$ is the spin of the droplet and $J$ is the ferromagnetic exchange constant between droplets. We further argue that most likely $S=1$ with four electrons occupying each droplet on average. We discuss the dependence of the magnetic susceptibility and the specific heat on the external magnetic field, which follows from the model, and hence we suggest further experimental tests of the model.

\end{abstract}

\date{\today}



\maketitle


\section{Introduction}

Studies of low density two-dimensional electron gas (2DEG) systems attract great attention, because of the unusual and rich properties of these systems. In the present work, we consider the magnetic properties of a silicon based 2DEG discovered in recent studies \cite{Reznikov_09,Reznikov_10,Reznikov_12}. Most likely the properties are related to the metal-insulator phase transition (MIT), and, in our opinion, understanding of these properties sheds some light on the nature of the transition.

The MIT in a 2DEG has attracted protracted attention from both experiment and theory, and remains a puzzling area of research to date \cite{Imada_98_MIT,Abrahams_01,Kravch_MIT-Review_04,Dobro_MIT-Review_11,Abrahams_79,Finkelstein_83,Finkelstein_84,Finkelstein_05,Kravch_95_big,Dobro_97_big,Kravch_97,Kravch_97b,Popovic_97,Kravch_98,Neilson_99,Vitkalov_01,Kravch_01}. It was once believed that a MIT in such systems could not take place, because a true metallic phase does not exist in a non-interacting 2DEG \cite{Abrahams_79}, although extension of the scaling theory of localisation to include the effects of interaction \cite{Finkelstein_83,Finkelstein_84} suggested that a MIT may be possible. Resistivity measurements in a silicon-based 2DEG finally provided evidence for a true MIT \cite{Kravch_95_big}, with numerous works following thereafter and studying the transition \cite{Dobro_97_big,Kravch_97,Kravch_97b,Popovic_97,Kravch_98,Neilson_99,Vitkalov_01,Kravch_01,Pudalov_02,Tutuc_02,Prus_03,Zhu_03}. The mechanism for a MIT in a 2DEG remains unclear to this date, but the existence of localised states on the insulator side of a MIT is generally accepted \cite{Mott,Anderson,Abrahams_01,Kravch_MIT-Review_04,Hamilton_PB_01}.

Intimately linked to the problem of the MIT is the nature of the ground state of a 2DEG, which still remains an outstanding problem \cite{Kravch_01,Tutuc_02,Prus_03,Zhu_03}. The ground state depends on electron density. There is little doubt that at a sufficiently high density it is a normal paramagnetic Fermi liquid \cite{Landau_all}. At a lower density, the system might have a Stoner transition to a ferromagnetic Fermi liquid \cite{Stoner}, and ultimately at a very low density it must undergo a transition to the Wigner crystal \cite{Wigner}. These are scenarios for a 2DEG without any extrinsic disorder, see Refs. \cite{Tanatar_89,Senatore_01,Bernu_01,Atta_02}. Extrinsic disorder can further complicate the situation, see, for instance, Refs. \cite{Voelker_01,Benenti_01}.

The magnetisation and magnetic susceptibility of a silicon based 2DEG in the electron density range below and above a MIT have been studied recently \cite{Prus_03,Reznikov_09,Reznikov_10,Reznikov_12}. The experiments have been performed with an in-plane magnetic field and so only spin related magnetic properties have been measured. There are three important outcomes of these measurements. (i) Thermodynamic magnetic properties vary continuously across the MIT. (ii) The zero-field magnetic susceptibility diverges rapidly in the limit $T \to 0$, $\chi_{0}\propto 1/T^{2.4}$, in both the insulating and metallic phases. (iii) In the metallic phase at $T \sim 1$ K, the value of the susceptibility is by orders of magnitude larger than the expected value of the ideal gas Pauli susceptibility.

The authors of Refs. \cite{Reznikov_09,Reznikov_10,Reznikov_12} explain their data through the formation of electron droplets. Each droplet has a nonzero spin. These droplets melt in the metallic phase with increasing density and temperature, but continue to exist up to large densities. At a fixed density of droplets, this picture would give the usual Curie scaling of the susceptibility with temperature, $\propto 1/T$. To explain the observed scaling, $\propto 1/T^{2.4}$, Ref. \cite{Reznikov_12} suggests that the density of droplets is decreasing when temperature is increasing.

In the present work, we take a somewhat different view to explain the magnetic data. We agree that the data practically unambiguously indicate the formation of electron droplets with nonzero spin. 
The droplets are probably formed due to extrinsic disorder or extrinsic
disorder assisted by the Coulomb interaction. Pragmatically for our purposes, the exact mechanism of their formation is not important. We only assume that at low temperatures all the internal degrees of freedom of the droplets are frozen and hence the only dynamical degree of freedom is the spin of the droplet.
The very steep temperature dependence of the magnetic susceptibility, in our opinion, indicates ferromagnetic instability. So we assume a ferromagnetic interaction between droplets and hence consider the system as a quantum Heisenberg model on a random 2D lattice. This is a very natural model for the insulating phase and we also assume that the localised droplets described by the Heisenberg model continue to exist in the metallic phase. While in the metallic phase most electrons go into itinerant states, these electrons are magnetically almost idle and the main contribution to the magnetic susceptibility is due to the relatively small fraction of electrons localised in the droplets. The density of the droplets diminishes in the metallic phase with increasing electron density.

The structure of the paper is as follows. In Section II, we review known properties of the 2D quantum Heisenberg ferromagnet and derive some previously unknown properties, which we need for the present work. A special point of the 2D Heisenberg ferromagnet is exponential dependence of the susceptibility on a special combination of spin and effective exchange constant. Comparison with experimental data performed in Section III allows us to determine this combination rather accurately. The number of electrons per droplet and the spin of the droplet appear only in a prefactor to the susceptibility. Therefore, we provide additional arguments to determine their values. In Section IV, we make predictions, which can be checked experimentally. Finally, Section V presents our conclusions.


\section{2D Quantum Heisenberg Ferromagnet}
The Heisenberg model is defined by the Hamiltonian
\begin{equation}
\label{H}
H_J=-J\sum_{<ij>}{\bm S}_i\cdot {\bm S}_j \ .
\end{equation}
We assume that the model is defined on a square lattice. Summation in (\ref{H}) is performed over nearest sites and ${\bm S}_i$ is the quantum spin at the site $i$. In the ground state, all spins are aligned ferromagnetically along, say, the z axis. Excitations are spin waves with the following spectrum \cite{Ashcroft-Mermin} 
\begin{equation}
\label{spin_fun2}
\varepsilon_{\textrm{k}} = 2 J S \left[2 - \cos(k_{\textrm{x}}) - \cos(k_{\textrm{y}}) \right] 
\xrightarrow[k \ll 1] ~JSk^2 \ .
\end{equation}
Hereafter we set the Planck constant and the Boltzmann constant equal to unity, $\hbar=k_B=1$. 

Each spinwave excitation carries spin $\Delta S_z=-1$. The excitations are bosons and therefore the magnetisation at a nonzero temperature is
\begin{equation}
\label{spin_fun1}
\left<S_{\textrm{z}}\right> = S - \int \frac{1}{e^{\varepsilon_{\textrm{k}}/T}-1} \frac{d^2 k}{(2\pi)^2} .
\end{equation}
The integral is logarithmically diverging at small momenta. This is a direct consequence of the Mermin-Wagner theorem \cite{Mermin-Wagner_66}, which claims that long range order is impossible in a 2D system at a nonzero temperature. The ferromagnetic ordering exists only within a correlation length $\xi$. To find value of $\xi$, one has to set $\left<S_{\textrm{z}}\right>=0$ and impose a lower limit in the integration in (\ref{spin_fun1}), $k > k_{min}\sim 1/\xi$. This gives the following correlation length in the low temperature limit \cite{Takahashi_87}
\begin{equation}
\label{spin_fun5}
\xi \propto e^{2\pi J S^2 / T} .
\end{equation}

There are $N \sim \xi^2$ spins within the correlation length, these spins act as a magnetic domain with total magnetic moment $M\sim S N$. The concentration of domains is $n_D\sim 1/N$. All in all, this describes a super-paramagnet with the following magnetic susceptibility, $\chi \propto n_DM^2 \propto N \propto \xi^2 \propto e^{4\pi J S^2 / T}$. This simple logic does not give the prefactor before the exponential. The renormalisation group (RG) calculation gives \cite{Kopietz89}
\begin{equation}
\label{chiRG}
\chi_{RG}=A\frac{S}{T}\left(\frac{T}{4\pi JS^2}\right)^3 e^{4\pi J S^2 / T} \ ,
\end{equation}
where $A$ is a constant. Note that the third power of the semiclassical parameter $\frac{T}{4\pi JS^2}$ in the prefactor in (\ref{chiRG}) arises in the two-loop approximation, while the single loop approximation gives only the first power of the parameter \cite{Kopietz89}.

\begin{figure}[h]
\includegraphics[clip=true,width=0.9\columnwidth]{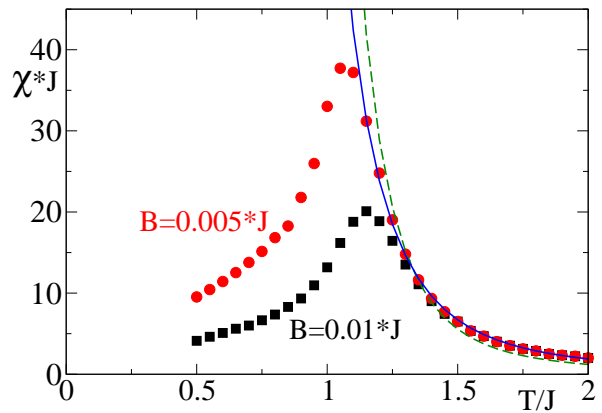}
\caption{Magnetic susceptibility per site obtained in QMC simulations of $S=1$ Heisenberg model\cite{Janke_08}. Values of the magnetic field are $B=0.005J$ (red circles) and $B=0.01J$ (black squares). The green dashed line shows fit to the RG Eq.~(\ref{chiRG}). The solid blue line shows the simple exponential fit (\ref{chifit1}).}
\label{MCdata}
\end{figure}

Quantum Monte Carlo (QMC) simulation of the magnetic susceptibility for $S=1/2$ was performed in Ref. \cite{Kopietz89} and for $S=1$ in Ref. \cite{Janke_08} In our analysis, we use the results of Ref. \cite{Janke_08}, because this simulation accounts for a nonzero magnetic field. The susceptibility QMC data \cite{Janke_08} for the values of magnetic field $B=0.005J$ and $B=0.01J$ are presented in Fig.~\ref{MCdata}.


Interaction with a magnetic field is defined by the Hamiltonian
\begin{equation}
\label{HB}
H_B=H_J -B\sum_i S_{iz} \ .
\end{equation}

Even a very small magnetic field significantly influences the susceptibility at low temperature. Eq.~(\ref{chiRG}) is valid only at low temperature, $JS > T$, but on the other hand, due to the presence of a magnetic field, the temperature cannot be too low, $T > T_m$. Here $T_m$ is the temperature where the susceptibility is maximum, see Fig.~\ref{MCdata}. The value of $T_m$ depends on the magnetic field. The very strong dependence of the susceptibility on the magnetic field is related to the dimensionality of the system. To explain the dependence, we remind that at zero temperature the magnon dispersion in a magnetic field at small k is\cite{Ashcroft-Mermin}
\begin{equation}
\label{SW_dispsn_finiteB}
\varepsilon_{\textrm{k}} = JS k^2 + B .
\end{equation}
When deriving Eqs.~(\ref{spin_fun5}) and (\ref{chiRG}), we substitute in (\ref{spin_fun1}) $k_{min}\sim 1/\xi$ as the lower limit of integration. This is correct only if $JS/\xi^2 > B$. From this condition, one immediately finds $T_m$
\begin{equation}
\label{Tm}
T_m \sim \frac{4\pi J S^2}{\ln\left(\frac{4\pi JS^2}{B} \right)} .
\end{equation}
The dependence on the magnetic field is logarithmic and hence even a tiny magnetic field significantly influences the susceptibility. The authors of Ref. \cite{Janke_08} used a power function to fit the dependence of $T_m$ on the magnetic field, $T_m \propto B^{\gamma}$. The fit gave a pretty small value $\gamma \approx 0.15$. While at $B \sim J$, the power fit might make sense, at small fields, $B \ll J$, the dependence is certainly logarithmic.

The RG expression (\ref{chiRG}) for magnetic susceptibility is valid at low temperature, $T \ll JS$, as it follows from the dispersion (\ref{spin_fun2}). On the other hand, the temperature must be higher than $T_m$. So the region of validity of Eq.~(\ref{chiRG}) is
\begin{equation}
\label{lm}
 T_m \ll T \ll JS  \ .
\end{equation}
Clearly, the right slope (green dashed line) in Fig.~\ref{MCdata} is not quite in this region, the magnetic field in these QMC data is not sufficiently small and hence the temperature is not sufficiently low. Nevertheless, we try to fit the data using Eq.~(\ref{chiRG}). The RG fit shown in Fig.~\ref{MCdata} by the green dashed line is not bad. However, the simple exponential fit
\begin{equation}
\label{chifit1}
\chi=0.042 S e^{7.6 J S^2 / T} \ ,
\end{equation}
shown by the blue solid line is better. Below in the analysis of experimental data we will use the simple exponential fit (\ref{chifit1}), having in mind that the data are also taken at not very low temperatures. Practically, this means that we will compare the experimental data directly with results of QMC simulations. Note that Eq.~(\ref{chifit1}) at $T \gg JS^2$ is approaching a constant instead of Curie's law, $\chi=S(S+1)/(3T)$. Hence Eq.~(\ref{chifit1}) overestimates the susceptibility at large $T$. In Eq.~(\ref{chifit1}), we assume linear scaling of the prefactor with $S$ as it is predicted by RG approach, see Eq.~(\ref{chiRG}). This is the scaling at a fixed value of the semiclassical parameter $JS^2/T$.

The electron spin droplets in silicon certainly do not form a regular square lattice. However, in the temperature range $4\pi J S^2  \gg T > T_m$, the exact structure of the lattice is not important. 
The dynamics depend only on the quadratic magnon dispersion, 
$\varepsilon_{\textrm{k}} = JS k^2$, where $J$ is an effective exchange constant.
The dispersion is the universal property of a 2D ferromagnet. 
Therefore, the exponential temperature dependence of the susceptibility is the universal property. Certainly one cannot seriously rely on the numerical prefactor, the prefactor is not universal. Eq.~(\ref{chifit1}) gives the magnetic susceptibility per lattice site. To compare with experimental data, we need to rewrite it per unit area and account for the Bohr magneton $\mu_B$ and $g$-factor of electrons in silicon, $g=2$. This gives in units of $\mu_B$ per tesla per $\textrm{cm}^2$ 
\begin{equation}
\label{chifit3}
\chi = 0.11 \frac{n_lS}{NJ}  e^{7.6 J S^2 / T} \ ,
\end{equation}
where $J$ is taken in Kelvin, $n_l$ is the density of localised electrons per $\textrm{cm}^2$ and $N$ is the number of electrons per droplet.

The low temperature ($T \ll JS$) and zero magnetic field specific heat per lattice site immediately follows from the dispersion (\ref{spin_fun2})
\begin{equation}
\label{sh}
C=\frac{T}{2\pi J S}\int_0^{\infty}\frac{xdx}{e^x-1}=\frac{\pi}{12}\frac{T}{JS} \ .
\end{equation}
Hence per unit area the specific heat is
\begin{equation}
\label{sh1}
C=\frac{\pi}{12}\frac{n_l}{N}\frac{T}{JS} \ .
\end{equation}
Interestingly, the temperature dependence is the same as that for a 2D Fermi gas, $C_F=\frac{\pi^2}{3}\frac{T}{\epsilon_F}$, however, the behaviour in an external magnetic field is distinctly different from that of a gas. The specific heat of the ferromagnet is suppressed in the field $B\sim T$, while the specific heat of the Fermi gas is not very sensitive to the field.

\section{Comparison with experimental magnetic susceptibility data}
\begin{figure}[h!]
\includegraphics[clip=true,width=0.8\columnwidth]{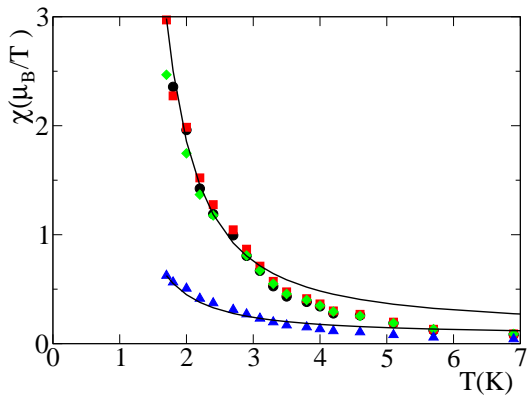}
\caption{Experimental data \cite{Reznikov_09} for $\chi/n$ versus temperature for various values of electron density $n$. Below values of $n$ are given in units $10^{11}/\textrm{cm}^2$. Black circles correspond to $n=0.55$, red squares correspond to $n=0.88$, green diamonds correspond to $n=1.43$ and blue triangles correspond to $n=4.3$. Lines show the exponential fits described in the text.}
\label{EXPdata}
\end{figure}

Experimental susceptibility data \cite{Reznikov_09} are presented in Fig.~\ref{EXPdata}. 
The MIT point occurs at $n = n_{\textrm{MIT}} = 0.85$ (Hereafter for densities we use the units $10^{11}$~cm$^{-2}$. Fit of experimental data by Eq.~(\ref{chifit3}) gives  $JS^2\approx 0.6 ~\textrm{K}$ for $n=0.55$, $0.88$ and $1.43$, while for $n=4.3$ 
it gives $JS^2\approx 0.4 ~\textrm{K}$. The fitted curves are shown in Fig.~\ref{EXPdata} by solid lines. For $T < 3~\textrm{K}$, the fits are very good, while at higher temperatures the fitted lines are above experimental points. There are two reasons for this. (i) We have pointed out above that at higher temperatures Eq.~(\ref{chifit3}) overestimates the susceptibility. (ii) It is likely that some electron droplets are thermally depopulated with rising temperature, reducing the susceptibility.

The above analysis, in our opinion, is very reliable and simple because it is based on a very steep exponential
dependence. After fixing the exponent $JS^2$, we discuss now the prefactor in Eq.~(\ref{chifit3}) and here we need to
attract additional arguments.
At the MIT and lower densities, all electrons are localised, hence $n_l=n$. Therefore, according to Eq.~(\ref{chifit3}), $\chi/n(T=1.7~\textrm{K}) \sim 1.6 S^3/N$. This should be compared with the experimental value $\chi/n(T=1.7~\textrm{K}) \approx 3$ presented in Fig.~\ref{EXPdata}. Very importantly,
the prefactor is of the right order of magnitude if $S\sim N \sim 1$. For $S=1/2$ and $N=1$, the prefactor is somewhat small compared to experiment, but still acceptable having in mind that theoretically the prefactor in Eq.~(\ref{chifit3}) is not very precise, so a disagreement by factor of several is quite possible. Other combinations with $S\sim N \sim 1$ are also acceptable from this point of view. In our opinion, the combination $S=1$ and $N=4$ is the most likely. There are the following arguments in favor of this combination. (i) The magnetic susceptibility prefactor is close to the experimental one. (ii) The experimental data of \cite{Prus_03,Reznikov_09} show that the saturation spin magnetisation in the insulating phase is roughly $\sim$50\% of the maximum possible spin magnetisation. For $S=1$ and $N=4$, one expects 50\% saturation magnetisation, which is close to the observed one. (iii) Many-body calculations for small 2D quantum dots \cite{Sushkov_05} show that the ground state spin of a four electron dot is $S=1$, already at a rather moderate value of Coulomb repulsion.
Note that in referring point (iii), we imply that the two valley dispersion degeneracy in 2D silicon is not relevant, the valley hybridisation is sufficiently large, at least 3-4 K. Due to the hybridisation, electrons occupy only the “bonding” combination of valleys and the droplet is similar to the $N = 4$ quantum dot in GaAs \cite{Sushkov_05}. Without such hybridisation, the spin of the $N = 4$ quantum dot in silicon would be $S = 0$, because two electrons with total spin $S = 0$ would occupy one valley and the other two electrons also with $S = 0$ would occupy the other valley. 
The $n=4.3$ curve in Fig.~\ref{EXPdata} lies significantly below the others. It is natural to assume that in the metallic phase ($n > n_{MIT}$), the number of localised electrons is roughly independent of $n$ and is equal to $n_l\approx n_{MIT}$. With account of this argument, the $n=4.3$ curve has to be scaled up by a factor of $4.3/0.85 \approx 5$. After this scaling, the curve is very close to the others.

All in all, the experimental data are remarkably consistent with the Heisenberg 
model picture of electron spin droplets.


\section{How to further check the Heisenberg model picture?}
The low temperature exponential divergence of the magnetic susceptibility in the 2D ferromagnetic Heisenberg model is necessarily accompanied by high sensitivity to the magnetic field. The dependence is clearly demonstrated in Fig.~\ref{MCdata}, where it manifests itself as $T_m(B)$ with two distinct behaviours for $T < T_m$  and $T > T_m$. Alternatively, one can consider the susceptibility (or magnetisation) as a function of the magnetic field at a fixed temperature. This will also have two distinct regimes, one with approximately linear dependence of the magnetisation on the magnetic field when $B < B^*$ and one with very slow increase/saturation of magnetisation at $B > B^*$. There are indications of such behaviour in existing experimental data \cite{Reznikov_09,Reznikov_10}. Further measurements and comparison with results of Monte Carlo simulations can shed more light on this problem.

Another possibility to test the model is to measure the specific heat in a magnetic field in the insulating phase. We already pointed out in Section II that the magnetic field significantly and predictably modifies the specific heat. Again, a comparison with the results of Monte Carlo simulations for the Heisenberg model would be very useful.

It is known that there are two mechanisms for interaction between localised spins: (i) usual exchange and (ii) superexchange. The exchange mechanism leads to the ferromagnetic interaction, while superexchange leads to the antiferromagnetic one. Let us denote by $R$ the radius of localisation (radius of the droplet) and by $l$ the separation between droplets. Obviously $R <l$.  Usually at $R \sim l$, exchange wins and this, in our opinion, describes the present situation. On the other hand, at $R \ll l$, superexchange always wins \cite{GP}, leading to the antiferromagnetic interaction. This implies that deeply in the insulating phase when $R \ll L$ we expect a transition to the antiferromagnetic interaction. Since the droplet positions are random, the antiferromagnetic interaction implies a spin glass state. So, we predict a transition to a spin glass state at a sufficiently low density and this must be clearly seen in the magnetic susceptibility. Unfortunately we cannot quantitatively predict the critical density for the onset of the spin glass state.


\section{Conclusions}

We suggest the quantum Heisenberg ferromagnet model to explain the anomalous magnetic properties observed in the vicinity of the metal insulator transition in a low-density two-dimensional silicon-based electron gas.  The ferromagnet is composed of electron spin droplets. The observed very steep temperature dependence of the magnetic susceptibility is associated with the exponential divergence in the Heisenberg model. By comparing the experimental data with known analytical and numerical results for the 2D Heisenberg ferromagnet, we determine the parameters of the model $JS^2 \approx 0.6~\textrm{K}$, where $S$ is the spin of the droplet and $J$ is the ferromagnetic exchange constant between droplets. 
We further argue that most likely $S=1$ with four electrons occupying each droplet on average. The 2D Heisenberg ferromagnet is strongly and distinctly influenced by magnetic field. Based on these properties, we suggest further experiments to test the model.


\section*{ACKNOWLEDGEMENTS}

We are grateful to M. Reznikov, A. R. Hamilton, J. Oitmaa, and C. J. Hamer for important discussions. We thank N. Teneh for communicating experimental data.

\end{document}